# Carrier and Spin Coherent Dynamics in Strained Germanium-Tin Semiconductor on Silicon

Sebastiano De Cesari[1], Andrea Balocchi[2, *], Elisa Vitiello[1], Pedram Jahandar[3], Emanuele Grilli[1], Thierry Amand[2], Xavier Marie[2], Maksym Myronov[3, *], Fabio Pezzoli[1, *].

[1]LNESS and Dipartimento di Scienza dei Materiali, Università degli Studi di Milano-Bicocca, via R. Cozzi 55, I-20125 Milan, Italy.

[2]Université Toulouse, INSA CNRS UPS, LPCNO, 135 Avenue Rangueil, F-31077 Toulouse, France.

[3]Department of Physics, The University of Warwick, Coventry CV4 7AL, United Kingdom.

*Corresponding authors: andrea.balocchi@insa-toulouse.fr, m.myronov@warwick.ac.uk, fabio.pezzoli@unimib.it.

**Germanium-Tin is emerging as a material exhibiting excellent photonic properties. Here we demonstrate optical initialization and readout of spins in this intriguing group IV semiconductor alloy and report on spin quantum beats between Zeeman-split levels under an external magnetic field. Our optical experiments reveal robust spin orientation in a wide temperature range and a persistent spin lifetime that approaches the ns regime at room temperature. Besides important insights into nonradiative recombination pathways, our findings disclose a rich spin physics in novel epitaxial structures directly grown on a conventional Si substrate. This introduces a viable route towards the synergic enrichment of the group IV semiconductor toolbox with advanced spintronics and photonic capabilities.**

The outstanding challenge of overcoming fundamental limits of conventional device electronics has stimulated various proposals and extensive investigations of radical alternatives[1]. The prospect of utilizing quantum information and communication processing has placed group IV semiconductors at the leading edge of current research efforts[2-4]. Such materials are ubiquitous in the mainstream microelectronic industry and naturally exhibit favourable properties for the solid-state implementation of logic-gate operations built upon quantum states[5]. The centrosymmetric crystal structure and the essential abundance of spin-less isotopes endow group IV semiconductors with long-lived electron spins[2, 6, 7] exhibiting exceedingly long coherence times[8-10].

Vast progress has been already achieved in the direction of quantum computation thanks to the demonstrations of spin-based quantum bits (qubits) using donor-bound electrons[11], valence band holes[12, 13] and gate-defined quantum dots[14-16]. Almost at the same time, the nontrivial s-p gap inversion occurring in the energy band-structure of the heavy element tin (Sn) has led to the discovery of exotic quantum phases of matter in group IV materials[17]. Besides the emergence of topological surface states[18, 19], strained layer epitaxy is tied to the remarkable modification of the topological invariant $Z_2$. Upon deposition on a lattice mismatched substrate, the nonmetallic allotrope, α-Sn, can be indeed converted from a $Z_2 = 0$ trivial insulator into either a nonzero $Z_2$ topological insulator or Dirac semimetal, i.e., the three dimensional analog of graphene[20]. Such findings potentially pave the way for advanced spintronic devices[21] and for a revolutionary realization of non-Abelian quantum computation in a Si compatible platform[22].

Hybridizing the extraordinary properties of α-Sn with the unique electronic states offered by another group IV element, namely Ge[6, 23], is promising for the tunability of quantum phases[24] and spin-dependent phenomena[7, 25-27]. To date, $Ge_{1-x}Sn_x$ binary alloys stand out among group IV materials for their superior optical properties[28-30] and transport mobilities[31-34]. Band-gap engineering[35, 36], enabled by very recent advances in crystal growth techniques[37-40], yield the fabrication of Ge-rich $Ge_{1-x}Sn_x$ epitaxial layers with a direct band-gap structure, simultaneously offering the key advantage of large-scale integration on Si wafers[30, 41].

The flexible design introduced by strain and alloying in terms of spin-orbit interaction can lead to optimal coupling between spin states and light fields, thus making semiconducting diluted-tin alloys a prominent candidate for bridging the gap between spintronic and photonic technologies[42]. To date, however, the spin properties of $Ge_{1-x}Sn_x$ remain completely unexplored. Moreover, the investigation of carrier recombination mechanisms is also at an early stage. Indeed, the study of the optical properties of $Ge_{1-x}Sn_x$ has been largely jeopardized by the demanding constraint of high-quality epitaxy. It has been hindered, above all, by fast and subtle nonradiative kinetic processes[43, 44] alongside the often-impractical long-wavelength regime associated to the optical transitions corresponding to the narrow band-gap. Consequently, there is a complete lack of crucial data including carrier and spin lifetimes.

Here we overcome the aforementioned limitations in an effort to address the spin dynamics and to assess the potential of $Ge_{1-x}Sn_x$ for the development of hybrid spin-based photonic concepts. To achieve this purpose, we focus on an indirect-gap $Ge_{0.95}Sn_{0.05}$ epitaxial layer, grown under compressive strain on a Ge buffered Si substrate via state-of-the-art chemical vapor deposition (see Fig. 1a). By doing so, we have access to a spectral window in the short-wave infrared that guarantees a reliable all-optical investigation of the kinetics of spin-polarized carriers. Thanks to pseudomorphic growth condition, we specifically inhibit the nucleation of detrimental crystal defects, mainly dislocations[43]. This is corroborated by an extensive structural characterization carried out throughout a broad array of techniques, including high-resolution X-ray diffraction, transmission electron and atomic force microscopy. See Fig. 1 and Methods for a detailed description. Our approach minimizes the occurrence of nonradiative channels in the optically-active compressive-strained $Ge_{1-x}Sn_x$ epilayer, and provides us with the possibility to better scrutinize intrinsic recombination mechanisms. Time- and polarization-resolved photoluminescence (PL) experiments reveal long-lived electron spin polarisation up to room temperature and allow us to identify the existence of recombination traps. Finally, spin quantum beat spectroscopy[45, 46] is revealed for the first time in group IV semiconductors by using PL, thus unveiling an effective electron g-factor of about 1.5. Our findings show the potential of novel Sn-based alloys to generate hybrid devices for quantum-information processing that simultaneously leverage spin and photon degrees of freedom.

**Optical features of pseudomorphic $Ge_{1-x}Sn_x$**

The low-temperature PL measured under continuous-wave excitation of a laser emitting at 1.165 eV is shown in the lower panel of Fig. 2a. The spectrum is characterized by a peak at about 0.64 eV, which is attributed to band-to-band recombination in the indirect-gap $Ge_{0.95}Sn_{0.05}$ epilayer involving *L*-valley electrons and heavy holes (see the sketch of the band structure in Fig. 2b). Owning to the Sn molar fraction and compressive strain, the direct gap is expected to lie ~100 meV higher in energy than the indirect gap[30, 43], while the strain-induced splitting between heavy and light hole bands is of about 60 meV[47]. The false-colour map of the PL intensity, presented in the upper panel of Fig. 2a, demonstrates that the increase of the lattice temperature yields a spectral broadening and a decrease of the PL amplitude. As pointed out very recently[43], the monotonic thermal-induced PL quenching, characterized in the lower panel of Fig. 2a (see the inset) by an activation energy of about 16 meV, is a compelling spectroscopic fingerprint of the pseudomorphic nature of the $Ge_{1-x}Sn_x$ ultrathin layer. This distinct PL behavior would have been otherwise concealed by the presence of nonradiative recombination channels opened up by dislocations. Such extended defects are introduced in the $Ge_{1-x}Sn_x$ epitaxial layers through plastic strain relaxation and are known to trigger a Schön-Klasens-like recombination dynamics, which, as opposed to the present case, enhances the PL intensity in the high temperature regime[43, 48-50].

Besides corroborating the structural characterization of the sample, the distinguished optical properties of the pseudomorphic layer allow us to perform optical orientation experiments[51-53] to verify the possibility of generating an out-of-equilibrium ensemble of spins directly in the $Ge_{1-x}Sn_x$ layer.

In Figure 2c we observe that, upon excitation with a $\sigma^+$ polarised laser, there is an imbalance between the right- ($I^{\sigma^+}$) and left- ($I^{\sigma^-}$) circularly polarised components of the PL. We also notice that the circular PL polarization degree, hereby defined as $\rho_{cir} = \frac{I^{\sigma^+} - I^{\sigma^-}}{I^{\sigma^+} + I^{\sigma^-}}$, is robust over a wide temperature range. Remarkably, $\rho_{cir}$ turns out to be sizeable even at room temperature and despite the off-resonance excitation condition of more than 500 meV. Such findings, along with the expected fast spin relaxation of holes[54, 55], unambiguously demonstrate the successful optical generation of a net spin polarisation of electrons in the conduction band (CB) of $Ge_{1-x}Sn_x$. Additionally, these noteworthy results clarify that the electron spin relaxation time is comparable or longer than the carrier lifetime.

So far we have considered the PL under steady state conditions. In the following, we apply optical spectroscopy in the time domain to explore the carrier kinetics and to gather a better understanding of the anticipated spin properties.

**Carrier kinetics in $Ge_{1-x}Sn_x$ binary alloys**

Figure 3 reports low-temperature time-resolved PL intensity measured at various pump fluencies. A non-exponential dynamics can be surprisingly appreciated. We start by considering the low excitation, i.e. $P_{AV}$ = 2.5 mW distributed over a spot diameter of ~7μm (black line in Fig. 3). Under such condition, the early phase of the recombination process exhibits a rather slow decay, which suddenly becomes steeper during the later stage of the temporal evolution (after ~4 ns). This puzzling PL transient can be accounted for by considering two intertwined recombination mechanisms. The fastest component of the decay curve suggests a large recombination probability controlled by an efficient nonradiative recombination channel. Its influence on the carrier kinetics is however masked during the slower initial decay phase. Knowing that the PL bleaching mimics the time evolution of the photo-excited population, we can conclude that the likely culprit for the initially diminished decay rate is the density of the excess carriers. At first, the pump pulse generates a large density of carriers that quite quickly saturates the low-energy states available for the fast recombination process. This leads to the spontaneous emergence of a shallower but less efficient recombination pathway characterized by a slower characteristic time. Later, the depopulation of the nonradiative centres activates the competing channel, whose effectiveness drastically accelerates the temporal decay of the PL.

The accuracy of this interpretation is further corroborated by the excitation intensity curves of Figure 3, which shows that a decreased pump power advances the change in slope of the decay curve, thus making the slow component less relevant for reduced densities of the photo-excited carriers.

It is worth noting that the same behavior can be consistently observed by increasing the lattice temperature (see Supplementary Fig. 1). Fig 3b summarizes the carrier lifetimes derived from the two slopes of the PL decay curves measured at various temperatures. In this case, we have carried out the measurements at a fixed intermediate pump power, i.e. $P_{AV}$ = 10 mW, in order to be able to better appreciate the two decay components over a large temperature range. Our experiments unveil that the carrier dynamics occurs within a couple of ns and is faster in the high temperature regime.

Since the band-to-band recombination pertaining to indirect gap materials develops on a slower time scale[7, 48], we attribute the measured temperature dependence of the lifetime to the presence of extrinsic recombination channels defined by shallow (~13-17 meV) energy levels (see Supplementary Note 1 for a detailed discussion). Such scenario compares satisfactorily with the temperature-induced quenching of the PL obtained under steady state conditions and summarized in Fig. 2a.

**Dependence of spin relaxation time upon carrier density and temperature**

In the following, we exploit time- and polarization-resolved PL experiments to gain a better knowledge of the spin-dependent properties and particularly of the spin dynamics. Figure 4a unambiguously demonstrates different intensities for the helicity-resolved PL transient following a circularly polarised excitation laser. This imbalance further corroborates the successful achievement of optical spin orientation of electrons in the $Ge_{1-x}Sn_x$ CB. In Fig. 4a, it can also be noted that the two polarization-resolved decay curves are almost parallel, thus indicating a rather long spin relaxation time, which is ultimately expected in the low power limit to be governed by the Elliot-Yafet mechanism[23].

The spin relaxation time ($T_1$) can be directly obtained by the decay time of $\rho_{cir}$ derived from the PL data. Figure 4b reveals that the pump drastically affects the spin dynamics, since the decay rate of the polarization degree strikingly increases with the laser power density. A variation in average excitation power from 2.5 to 30 mW shortens the spin relaxation time from about 60 to 10 ns (see the inset of Fig. 4b). Such measurements indicate the occurrence of optically-induced spin-flip processes, thus suggesting the importance of electron-hole Coulomb exchange interaction amongst the out-of-equilibrium spin ensemble[6, 51, 56].

This finding is further supported by a systematic investigation of the spin lifetime as a function of temperature. As clarified in the Supplementary Fig. 3, the experimental values of $T_1$ have been accurately tested by a direct comparison with the temperature-dependent characteristics of the carrier lifetime (Fig. 3b) and the circular polarization degree of the PL measured under steady state conditions (Fig. 2c).

Fig 4c shows that $T_1$ steadily decreases as the temperature is increased from 7 to 200 K. Interestingly, $T_1$ remains rather constant at higher temperatures. Such behavior mimics the one of the carrier lifetime shown in Fig. 3b and suggests their possible entangled origin. The following scenario likely accounts for such observations. At low temperatures, the electron spin relaxation is dominated, under our experimental conditions, by exchange interaction with photoinduced holes. In this temperature regime, the optical activity of the defect states, being largely occupied by extrinsic carriers, turns out to be mitigated. Thermal-induced ionization of the shallow traps, however, activates nonradiative recombination events that shorten the carrier lifetime and eventually increases the concentration of unpolarized free carriers. The latter can contribute along with scattering off the central-cell potential of the impurities[57] and the phonon mediated spin-flip processes[23, 58] to the temperature-driven Elliot-Yaffet spin relaxation observed in Fig 4c.

It should be noted that $T_1$ remains in the ns regime even in the high temperature range and approaches the values theoretically predicted in pure Ge[23]. Such result suggests that the small amount of the heavier element Sn, incorporated in the lattice, gently perturbs the electron spin properties, thus preserving (i) an almost perfect Oh symmetry of the lattice, so that Dyakonov-Perel relaxation is negligible and (ii) the favorable spin-orbit coupling pertaining to the Ge band structure. The latter has been remarkably identified as a key feature that enables Ge to support exceedingly long spin lifetimes[6, 23].

In light of these considerations, we expect that alloying Ge with Sn can offer a novel degree of freedom to engineer phenomena based on spin-orbit coupling, such as Rashba[59] or spin Hall[26] effects, while retaining a desirably long spin relaxation[6]. Finally, we notice that the $T_1$ regime measured at room temperature is already comparable to the typical switching frequency of standard electronic devices, thus increasing the prospects of a practical implementation of $Ge_{1-x}Sn_x$ for spin transport and manipulation in future device architectures[60]. Such result, if combined with high light-matter interaction and the high optical quality offered by $Ge_{1-x}Sn_x$ binary alloys can be a key advantage in designing future spin-based optoelectronic devices on the Si platform.

**Quantum beat spectroscopy and coherent spin dynamics**

In the following, we extend our all-optical investigation by measuring the electron spin dynamics under the presence of an external magnetic field in Voigt geometry. In particular, we apply spin quantum beat spectroscopy (QBS) to Ge-based systems, demonstrating coherent spin dynamics and providing an estimation of the Landé $g$-factor of CB electrons in strained $Ge_{1-x}Sn_x$ layers.

The intensity oscillations of the PL circular polarisation displayed in Figure 5a reflect the Larmor precession of the electron spins in the transverse magnetic field. The precession frequency $\omega$ depends on the strength of the magnetic field $B$ as $\hbar\omega = g^*\mu_B B$, where $\mu_B$ is the Bohr's magneton and $g^*$ is the effective $g$-factor. Figure 5a and 5b nicely demonstrate such a linear relationship, and allow us to determine the effective Landé factor $g^*$ of the conduction electrons. We find $g^*$=1.48±0.01. This value compares favourably with recent reports of the Landé $g$-factor of $L$-valleys electrons in Ge[6, 61].

We observe in Figure 5 that the magnetic field amplitude affects the period of the quantum beats along with their damping time. The latter feature, summarized in Fig. 5c, discloses the presence of transverse spin relaxation mechanisms. The substantial decrease of the dephasing time of the spin ensemble ($T_2^*$) with the strength of applied magnetic field is a result of $g$-factor fluctuations ($\Delta g$)[62, 63]. Indeed, in the optical orientation process, light absorption generates CB electrons with a finite excess energy. Spin dephasing is triggered by the cooling process, during which electrons experience momentum relaxation and random effective magnetic fields given by local changes in the alloy composition and strain. Data in Fig. 5c are well described by the equation[63] $1/T_2^*(B) = 1/T_2^*(0) + \Delta g \mu_B B/\hbar$ (black line) resulting from a normal distribution of g-factors, and provides us with a $\Delta g$ of $2.1 \times 10^{-2}$.

## Discussion and conclusions

We have explored the exciting prospect of utilizing a new group IV material, namely $Ge_{1-x}Sn_x$, as future solid-state hosts of integrated quantum information and communication processing. The tunable band structure and spin-orbit coupling, promised by this novel alloy, can potentially offer a very rich spin physics, whose fundamental understanding is however still absent. The demonstrated applicability of light-matter interaction facilitated in this work the anticipation of spin-dependent phenomena pertaining to this notable material. Robust quantum coherence within a spin ensemble has been observed and loss mechanisms induced by local alloy fluctuations identified. Finally, the PL dynamics measurements allowed us to gather insights into recombination mechanisms and to resolve the physics of carrier and spin dynamics over a wide temperature range. These findings can be precious in the quest for achieving real-life deployment of light emitting devices based on $Ge_{1-x}Sn_x$ alloys. Looking ahead, the determination of the effective Landé g-factor and the temperature-dependent spin-flip mechanisms can provide key information for the future design of spintronic devices and the potential applicability of $Ge_{1-x}Sn_x$ for gate-defined spin-qubits and spin-photon interfaces in distributed quantum networks.

# Bibliography

1. Waldrop, M. M., The chips are down for Moore's law. *Nature* **2016,** *530*, 144-147.

2. Morton, J. J. L.; McCamey, D. R.; Eriksson, M. A.; Lyon, S. A., Embracing the quantum limit in silicon computing. *Nature* **2011,** *479*, 345-353.

3. Jansen, R., Silicon spintronics. *Nature Materials* **2012,** *11*, 400-408.

4. Morrison, C.; Myronov, M., Strained germanium for applications in spintronics. *Physica Status Solidi a-Applications and Materials Science* **2016,** *213*, 2809-2819.

5. Zwanenburg, F. A.; Dzurak, A. S.; Morello, A.; Simmons, M. Y.; Hollenberg, L. C. L.; Klimeck, G.; Rogge, S.; Coppersmith, S. N.; Eriksson, M. A., Silicon quantum electronics. *Reviews of Modern Physics* **2013,** *85*, 961-1019.

6. Giorgioni, A.; Paleari, S.; Cecchi, S.; Vitiello, E.; Grilli, E.; Isella, G.; Jantsch, W.; Fanciulli, M.; Pezzoli, F., Strong confinement-induced engineering of the g factor and lifetime of conduction electron spins in Ge quantum wells. *Nature Communications* **2016,** *7*, 13886.

7. Giorgioni, A.; Vitiello, E.; Grilli, E.; Guzzi, M.; Pezzoli, F., Valley-dependent spin polarization and long-lived electron spins in germanium. *Applied Physics Letters* **2014,** *105*, 152404.

8. Tyryshkin, A. M.; Tojo, S.; Morton, J. J. L.; Riemann, H.; Abrosimov, N. V.; Becker, P.; Pohl, H. J.; Schenkel, T.; Thewalt, M. L. W.; Itoh, K. M.; Lyon, S. A., Electron spin coherence exceeding seconds in high-purity silicon. *Nature Materials* **2012,** *11*, 143-147.

9. Sigillito, A. J.; Jock, R. M.; Tyryshkin, A. M.; Beeman, J. W.; Haller, E. E.; Itoh, K. M.; Lyon, S. A., Electron Spin Coherence of Shallow Donors in Natural and Isotopically Enriched Germanium. *Physical Review Letters* **2015,** *115*, 247601.

10. Muhonen, J. T.; Dehollain, J. P.; Laucht, A.; Hudson, F. E.; Kalra, R.; Sekiguchi, T.; Itoh, K. M.; Jamieson, D. N.; McCallum, J. C.; Dzurak, A. S.; Morello, A., Storing quantum information for 30 seconds in a nanoelectronic device. *Nature Nanotechnology* **2014,** *9*, 986-991.

11. Pla, J. J.; Tan, K. Y.; Dehollain, J. P.; Lim, W. H.; Morton, J. J. L.; Jamieson, D. N.; Dzurak, A. S.; Morello, A., A single-atom electron spin qubit in silicon. *Nature* **2012,** *489*, 541-545.

12. Maurand, R.; Jehl, X.; Kotekar-Patil, D.; Corna, A.; Bohuslavskyi, H.; Lavieville, R.; Hutin, L.; Barraud, S.; Vinet, M.; Sanquer, M.; De Franceschi, S., A CMOS silicon spin qubit. *Nature Communications* **2016,** *7*, 13575.

13. Hu, Y. J.; Kuemmeth, F.; Lieber, C. M.; Marcus, C. M., Hole spin relaxation in Ge-Si core-shell nanowire qubits. *Nature Nanotechnology* **2012,** *7*, 47-50.

14. Kawakami, E.; Scarlino, P.; Ward, D. R.; Braakman, F. R.; Savage, D. E.; Lagally, M. G.; Friesen, M.; Coppersmith, S. N.; Eriksson, M. A.; Vandersypen, L. M. K., Electrical control of a long-lived spin qubit in a Si/SiGe quantum dot. *Nature Nanotechnology* **2014,** *9*, 666-670.

15. Veldhorst, M.; Yang, C. H.; Hwang, J. C. C.; Huang, W.; Dehollain, J. P.; Muhonen, J. T.; Simmons, S.; Laucht, A.; Hudson, F. E.; Itoh, K. M.; Morello, A.; Dzurak, A. S., A two-qubit logic gate in silicon. *Nature* **2015,** *526*, 410-414.

16. Henneberger, F.; Benson, O., *Semiconductor quantum bits*. Singapore, Pan Stanford Publishing, 2009.

17. Fu, L.; Kane, C. L., Topological insulators with inversion symmetry. *Physical Review B* **2007,** *76*, 045302.

18. Barfuss, A.; Dudy, L.; Scholz, M. R.; Roth, H.; Hopfner, P.; Blumenstein, C.; Landolt, G.; Dil, J. H.; Plumb, N. C.; Radovic, M.; Bostwick, A.; Rotenberg, E.; Fleszar, A.; Bihlmayer, G.; Wortmann, D.; Li, G.; Hanke, W.; Claessen, R.; Schafer, J., Elemental Topological Insulator with Tunable Fermi Level: Strained alpha-Sn on InSb(001). *Physical Review Letters* **2013,** *111*, 157205.

19. Ohtsubo, Y.; Le Fevre, P.; Bertran, F.; Taleb-Ibrahimi, A., Dirac Cone with Helical Spin Polarization in Ultrathin alpha-Sn(001) Films. *Physical Review Letters* **2013,** *111*, 216401.

20. Xu, C. Z.; Chan, Y. H.; Chen, Y. G.; Chen, P.; Wang, X. X.; Dejoie, C.; Wong, M. H.; Hlevyack, J. A.; Ryu, H. J.; Kee, H. Y.; Tamura, N.; Chou, M. Y.; Hussain, Z.; Mo, S. K.; Chiang, T. C., Elemental Topological Dirac Semimetal: alpha-Sn on InSb(111). *Physical Review Letters* **2017,** *118*, 146402.

21. Han, J.; Richardella, A.; Siddiqui, S. A.; Finley, J.; Samarth, N.; Liu, L., Room-Temperature Spin-Orbit Torque Switching Induced by a Topological Insulator. *Physical Review Letters* **2017,** *119*, 077702.

22. Moore, J. E., The birth of topological insulators. *Nature* **2010,** *464*, 194-198.

23. Li, P.; Song, Y.; Dery, H., Intrinsic spin lifetime of conduction electrons in germanium. *Physical Review B* **2012,** *86,* 085202.

24. Lan, H. S.; Chang, S. T.; Liu, C. W., Semiconductor, topological semimetal, indirect semimetal, and topological Dirac semimetal phases of Ge1-xSnx alloys. *Physical Review B* **2017,** *95*, 201201(R).

25. Pezzoli, F.; Qing, L.; Giorgioni, A.; Isella, G.; Grilli, E.; Guzzi, M.; Dery, H., Spin and energy relaxation in germanium studied by spin-polarized direct-gap photoluminescence. *Physical Review B* **2013,** *88*, 045204.

26. Bottegoni, F.; Zucchetti, C.; Dal Conte, S.; Frigerio, J.; Carpene, E.; Vergnaud, C.; Jamet, M.; Isella, G.; Ciccacci, F.; Cerullo, G.; Finazzi, M., Spin-Hall Voltage over a Large Length Scale in Bulk Germanium. *Physical Review Letters* **2017,** *118,* 167402.

27. Pezzoli, F.; Balocchi, A.; Vitiello, E.; Amand, T.; Marie, X., Optical orientation of electron spins and valence-band spectroscopy in germanium. *Physical Review B* **2015,** *91,* 201201(R).

28. Soref, R. A.; Friedman, L., Direct-gap Ge/GeSn/Si and GeSn/Ge/Si heterostructures. *Superlattices and Microstructures* **1993,** *14*, 189.

29. Soref, R., Mid-infrared photonics in silicon and germanium. *Nature Photonics* **2010,** *4*, 495-497.

30. Wirths, S.; Geiger, R.; von den Driesch, N.; Mussler, G.; Stoica, T.; Mantl, S.; Ikonic, Z.; Luysberg, M.; Chiussi, S.; Hartmann, J. M.; Sigg, H.; Faist, J.; Buca, D.; Grutzmacher, D., Lasing in direct-bandgap GeSn alloy grown on Si. *Nature Photonics* **2015,** *9*, 88-92.

31. Kotlyar, R.; Avci, U. E.; Cea, S.; Rios, R.; Linton, T. D.; Kuhn, K. J.; Young, I. A., Bandgap engineering of group IV materials for complementary n and p tunneling field effect transistors. *Applied Physics Letters* **2013,** *102*, 113106.

32. Sau, J. D.; Cohen, M. L., Possibility of increased mobility in Ge-Sn alloy system. *Physical Review B* **2007,** *75*, 045208.

33. Gupta, S.; Chen, R.; Magyari-Kope, B.; Lin, H.; Yang, B.; Nainani, A.; Nishi, Y.; Harris, J. S.; Saraswat, K. C., GeSn Technology: Extending the Ge Electronics Roadmap. *2011 IEEE International Electron Devices Meeting (Iedm)* **2011**, 16.6.1-16.6.4.

34. Han, G. Q.; Su, S. J.; Zhan, C. L.; Zhou, Q.; Yang, Y.; Wang, L. X.; Guo, P. F.; Wei, W.; Wong, C. P.; Shen, Z. X.; Cheng, B. W.; Yeo, Y. C., High-Mobility Germanium-Tin (GeSn) P-channel MOSFETs Featuring Metallic Source/Drain and Sub-370 degrees C Process Modules. *2011 Ieee International Electron Devices Meeting (Iedm)* **2011**, 16.7.1-16.7.3.

35. Soref, R. A.; Perry, C. H., PREDICTED BAND-GAP OF THE NEW SEMICONDUCTOR SIGESN. *Journal of Applied Physics* **1991,** *69*, 539-541.

36. Gupta, S.; Magyari-Kope, B.; Nishi, Y.; Saraswat, K. C., Achieving direct band gap in germanium through integration of Sn alloying and external strain. *Journal of Applied Physics* **2013,** *113*, 073707.

37. D'Costa, V. R.; Fang, Y. Y.; Tolle, J.; Kouvetakis, J.; Menéndez, J., Tunable Optical Gap at a Fixed Lattice Constant in Group-IV Semiconductor Alloys. *Physical Review Letters* **2009,** *102*, 107403.

38. Wirths, S.; Buca, D.; Mantl, S., Si-Ge-Sn alloys: From growth to applications. *Progress in Crystal Growth and Characterization of Materials* **2016,** *62*, 1-39.

39. Vincent, B.; Gencarelli, F.; Bender, H.; Merckling, C.; Douhard, B.; Petersen, D. H.; Hansen, O.; Henrichsen, H. H.; Meersschaut, J.; Vandervorst, W.; Heyns, M.; Loo, R.; Caymax, M., Undoped and in-situ B doped GeSn epitaxial growth on Ge by atmospheric pressure-chemical vapor deposition. *Applied Physics Letters* **2011,** *99*, 152103.

40. Bauer, M.; Ritter, C.; Crozier, P. A.; Ren, J.; Menendez, J.; Wolf, G.; Kouvetakis, J., Synthesis of ternary SiGeSn semiconductors on Si(100) via SnxGe1-x buffer layers. *Applied Physics Letters* **2003,** *83*, 2163-2165.

41. Ghetmiri, S. A.; Du, W.; Margetis, J.; Mosleh, A.; Cousar, L.; Conley, B. R.; Domulevicz, L.; Nazzal, A.; Sun, G.; Soref, R. A.; Tolle, J.; Li, B. H.; Naseem, H. A.; Yu, S. Q., Direct-bandgap GeSn grown on silicon with 2230 nm photoluminescence. *Applied Physics Letters* **2014,** 151109.

42. De Cesari, S.; Vitiello, E.; Giorgioni, A.; Pezzoli, F., Progress towards Spin-Based Light Emission in Group IV Semiconductors. *Electronics* **2017,** *6*, 25.

43. Pezzoli, F.; Giorgioni, A.; Patchett, D.; Myronov, M., Temperature-Dependent Photoluminescence Characteristics of GeSn Epitaxial Layers. *Acs Photonics* **2016,** *3*, 2004-2009.

44. Gallagher, J. D.; Senaratne, C. L.; Sims, P.; Aoki, T.; Menendez, J.; Kouvetakis, J., Electroluminescence from GeSn heterostructure pin diodes at the indirect to direct transition. *Applied Physics Letters* **2015,** *106,* 091103.

45. Amand, T.; Marie, X.; LeJeune, P.; Brousseau, M.; Robart, D.; Barrau, J.; Planel, R., Spin quantum beats of 2D excitons. *Physical Review Letters* **1997,** *78*, 1355-1358.

46. Heberle, A. P.; Ruhle, W. W.; Ploog, K., QUANTUM BEATS OF ELECTRON LARMOR PRECESSION IN GAAS WELLS. *Physical Review Letters* **1994,** *72*, 3887-3890.

47. Zelazna, K.; Polak, M. P.; Scharoch, P.; Serafinczuk, J.; Gladysiewicz, M.; Misiewicz, J.; Dekoster, J.; Kudrawiec, R., Electronic band structure of compressively strained Ge1-xSnx with x < 0.11 studied by contactless electroreflectance. *Applied Physics Letters* **2015,** *106,* 142102.

48. Pezzoli, F.; Giorgioni, A.; Gallacher, K.; Isa, F.; Biagioni, P.; Millar, R. W.; Gatti, E.; Grilli, E.; Bonera, E.; Isella, G.; Paul, D. J.; Miglio, L., Disentangling nonradiative recombination processes in Ge micro-crystals on Si substrates. *Applied Physics Letters* **2016,** *108,* 262103.

49. Pezzoli, F.; Isa, F.; Isella, G.; Falub, C. V.; Kreiliger, T.; Salvalaglio, M.; Bergamaschini, R.; Grilli, E.; Guzzi, M.; von Kanel, H.; Miglio, L., Ge Crystals on Si Show Their Light. *Physical Review Applied* **2014,** *1*, 044005.

50. Reshchikov, M. A., Temperature dependence of defect-related photoluminescence in III-V and II-VI semiconductors. *Journal of Applied Physics* **2014,** *115*, 012010.

51. Dyakonov, M. I.; Perel, V. I., In *Optical Orientation*, Meier, F., Ed. North Holland, Amsterdam, 1984.

52. Lampel, G., Nuclear Dynamic Polarization by Optical Electronic Saturation and Optical Pumping in Semiconductors. *Physical Review Letters* **1968,** *20*, 491-493.

53. Zutic, I.; Fabian, J.; Das Sarma, S., Spintronics: Fundamentals and applications. *Reviews of Modern Physics* **2004,** *76*, 323-410.

54. Loren, E. J.; Rioux, J.; Lange, C.; Sipe, J. E.; van Driel, H. M.; Smirl, A. L., Hole spin relaxation and intervalley electron scattering in germanium. *Physical Review B* **2011,** *84,* 214307.

55. Lange, C.; Isella, G.; Chrastina, D.; Pezzoli, F.; Koester, N. S.; Woscholski, R.; Chatterjee, S., Spin band-gap renormalization and hole spin dynamics in Ge/SiGe quantum wells. *Physical Review B* **2012,** *85,* 241303(R).

56. Zerrouati, K.; Fabre, F.; Bacquet, G.; Bandet, J.; Frandon, J.; Lampel, G.; Paget, D., Spin-lattice relaxation in p-type gallium arsenide single crystals. *Physical Review B* **1988,** *37*, 1334-1341.

57. Song, Y.; Chalaev, O.; Dery, H., Donor-Driven Spin Relaxation in Multivalley Semiconductors. *Physical Review Letters* **2014,** *113*, 167201.

58. Tang, J.-M.; Collins, B. T.; Flatte, M. E., Electron spin-phonon interaction symmetries and tunable spin relaxation in silicon and germanium. *Physical Review B* **2012,** *85,* 045202.

59. Wilamowski, Z.; Jantsch, W.; Malissa, H.; Rossler, U., Evidence and evaluation of the Bychkov-Rashba effect in SiGe/Si/SiGe quantum wells. *Physical Review B* **2002,** *66,* 195315.

60. Dushenko, S.; Koike, M.; Ando, Y.; Shinjo, T.; Myronov, M.; Shiraishi, M., Experimental Demonstration of Room-Temperature Spin Transport in n-Type Germanium Epilayers. *Physical Review Letters* **2015,** *114,* 196602.

61. Hautmann, C.; Betz, M., Magneto-optical analysis of the effective g tensor and electron spin decoherence in the multivalley conduction band of bulk germanium. *Physical Review B* **2012,** *85*, 121203.

62. Margulis, A. D.; Margulis, V. A., PRECESSION MECHANISM OF CONDUCTIVE ELECTRON-SPIN RELAXATION IN SEMICONDUCTORS IN A STRONG MAGNETIC-FIELD. *Fizika Tverdogo Tela* **1983,** *25*, 1590-1596.

63. Belykh, V. V.; Greilich, A.; Yakovlev, D. R.; Yacob, M.; Reithmaier, J. P.; Benyoucef, M.; Bayer, M., Electron and hole g factors in InAs/InAlGaAs self-assembled quantum dots emitting at telecom wavelengths. *Physical Review B* **2015,** *92*, 165307.

## Acknowledgements

This work was supported by Fondazione CARIPLO through the project SEARCH-IV, Grant No. 2013.0623.

## Methods

**Sample growth.** The $Ge_{1-x}Sn_x$/Ge/Si(001) heterostructure was grown within a ASM Epsilon 2000 industrial type reduced pressure chemical vapour deposition (RP-CVD) system. The strained $Ge_{1-x}Sn_x$ epilayer was deposited on a 100 mm diameter Si (001) substrate via a relaxed Ge buffer with thickness ~650 nm. $SnCl_4$ was used as a Sn precursor while $Ge_2H_6$ was used as the Ge precursor. Growth was carried out at a temperature of 280 °C in a $H_2$ atmosphere.

**Structural characterization.** A Jeol JEM-2100 transmission electron microscope (TEM) was used to obtain high-resolution cross-sectional micrographs of the heterostructure (X-TEM). TEM imaging allows direct observation of the crystalline quality and measurements of epilayer's thickness. The surface morphology of the heterostructure was mapped using an Asylum Research MFP-3D stand-alone atomic force microscope (AFM). The surface topology were acquired using Silicon Nitride ($Si_3N_4$) tip on the AFM operating in a tapping mode. The root mean square roughness turned out to be < 2 nm. Finally, the Sn molar fraction of 5% and the biaxial compressive strain of 0.80% of the $Ge_{1-x}Sn_x$ epilayer were measured from symmetrical (004) and asymmetrical (224) high-resolution X-ray diffraction (HR-XRD) reciprocal space maps. This was carried out via $\omega$-2$\theta$ coupled scans and symmetric and asymmetric reciprocal space maps (RSMs) on a Panalytical X'Pert Pro MRD using $CuK\alpha_1$ source.

**Optical investigation.** Continuous-wave photoluminescence (PL) experiments were performed by exciting the sample with a Nd-$YVO_4$ laser at 1.165 eV. The spot diameter was of about 50 μm, and the resulting power density was of few kW/cm$^2$. The polarized PL was at first selected by combining a $\lambda/4$ waveplate with a linear polarizer, and then dispersed a monochromator equipped with an InGaAs detector with a cut-off at ~0.5 eV. The spectra were numerically cleaned to remove the overlap with the second order peak of the pump. Time-resolved PL was carried out by using the excitation energy of 1.165 eV delivered by an optical parameter oscillator pumped by a Ti:Sa laser.

The temporal width and the repetition rate of the laser pulses were of ~1 ps and 80 MHz, respectively. The laser spot size was estimated to be of about 7 μm and the average excitation power was varied from 2.5 mW to 30 mW. The PL was spectrally selected by means of band pass filters and the polarization was resolved via a quarter-waveplate and a linear polarizer. Finally, the PL dynamics was determined by applying time-correlated single-photon counting using a detector with a long-wavelength cut-off and time resolution of either 64 or 128 ps.

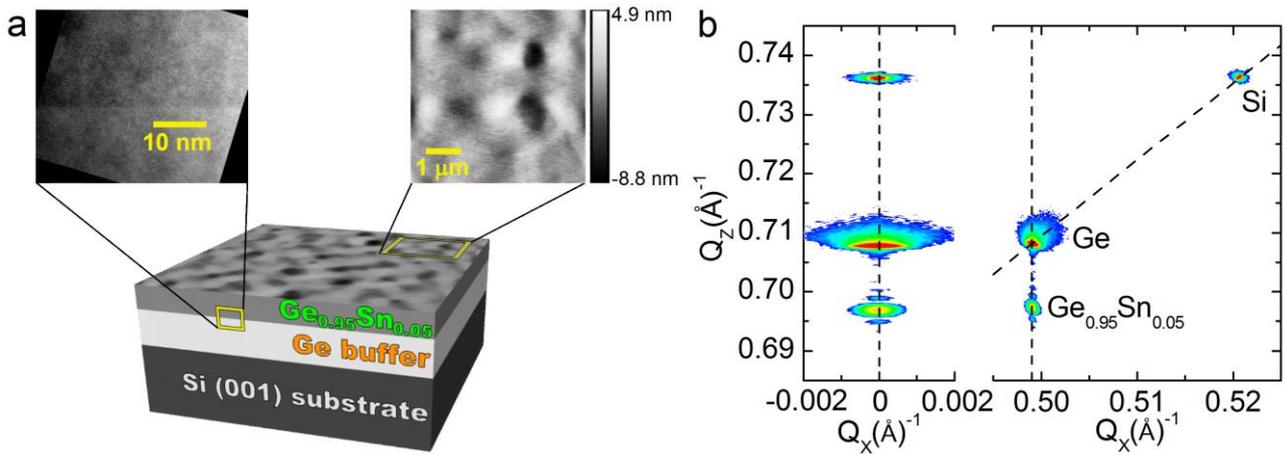

**Figure 1| Structural properties of the GeSn-based heterostructure. a** Sketch of the $Ge_{0.95}Sn_{0.05}$/Ge/Si heterostructure (not to scale). Cross-sectional Transmission Electron Microscope image highlighting the GeSn/Ge hetero-interface (left panel) and Atomic Force Microscopy micrograph of the surface morphology of the sample (right panel). **b** X-ray diffraction (004) symmetric (left) and (224) asymmetric (right) reciprocal space maps. In the latter, the Bragg spots pertaining to Ge and GeSn are precisely aligned along the diffraction wave vector, $Q_x$, spanning the sample surface. This explicitly demonstrates the attained coherent growth condition. The Sn-based epilayer is indeed in perfect in-plan registry with the underlying Ge buffer.

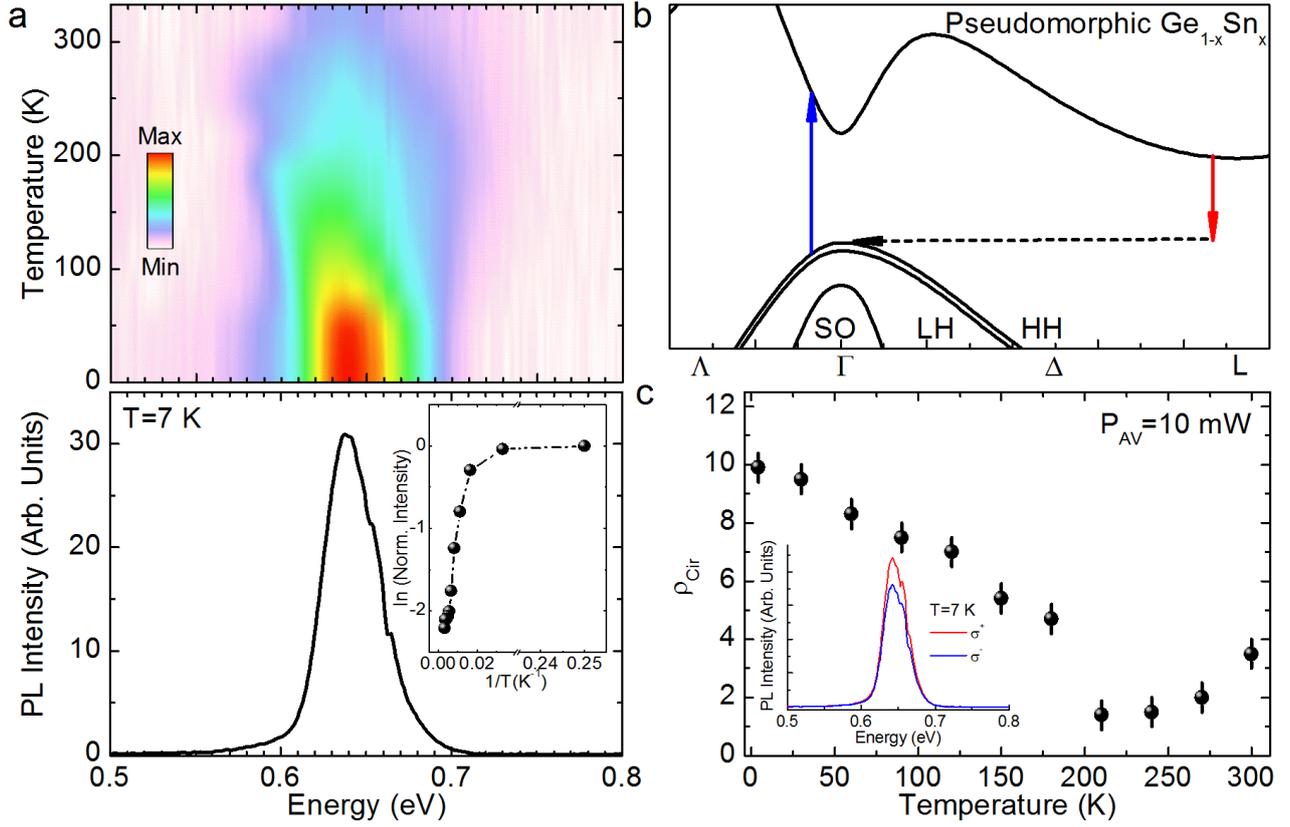

**Figure 2| Optical properties of the GeSn-based heterostructure. a** Upper panel: Colour-coded photoluminescence (PL) intensity as a function of the lattice temperature for the $Ge_{0.95}Sn_{0.05}$ epitaxial layer. Lower panel: Low temperature PL spectrum. The inset shows the temperature dependence of the spectrally integrated intensity. **b** Schematic representation of the band structure pertaining to a pseudomorphic $Ge_{0.95}Sn_{0.05}$ layer grown under compressive strain on a Ge-buffered Si substrate. Arrows show the absorption and emission processes though the direct and indirect gap, respectively. **c** Temperature dependence of the circular polarization degree, $\rho_{circ}$, defined as $\frac{I^{\sigma^+} - I^{\sigma^-}}{I^{\sigma^+} + I^{\sigma^-}}$. Here $I^{\sigma^+}$ and $I^{\sigma^-}$ are the PL circular components. The inset shows the PL spectra for the $Ge_{0.95}Sn_{0.05}$ film resolved for right- and left-handed circular polarizations, namely $\sigma^+$ and $\sigma^-$ respectively. The spectra have been obtained a temperature of 4K by utilizing a $\sigma^+$ excitation energy at 1.165 eV.

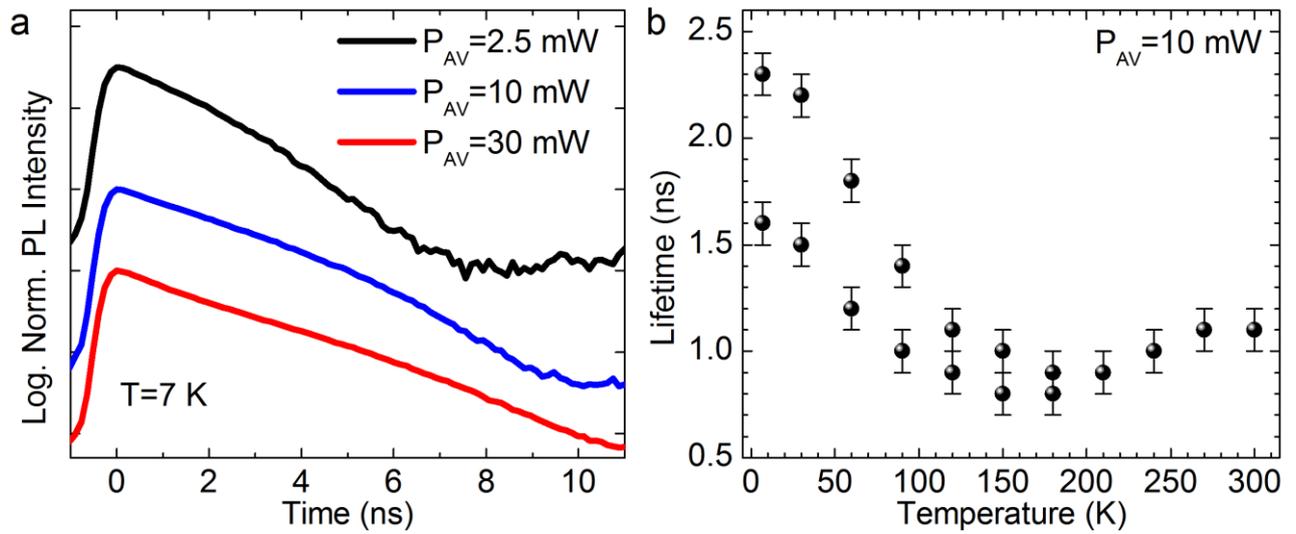

**Figure 3| Time-resolved photoluminescence. a** Low-temperature normalised PL dynamics of the pseudomorphic $Ge_{0.95}Sn_{0.05}$ epitaxial layer as a function of the average excitation power. The pump pulse has an energy of ~1.165 eV and a duration of about 2 ps. **b** Dependence of the carrier recombination time on the lattice temperature. All the measurements have been carried out at an average excitation power of $P_{AV} = 10$ mW.

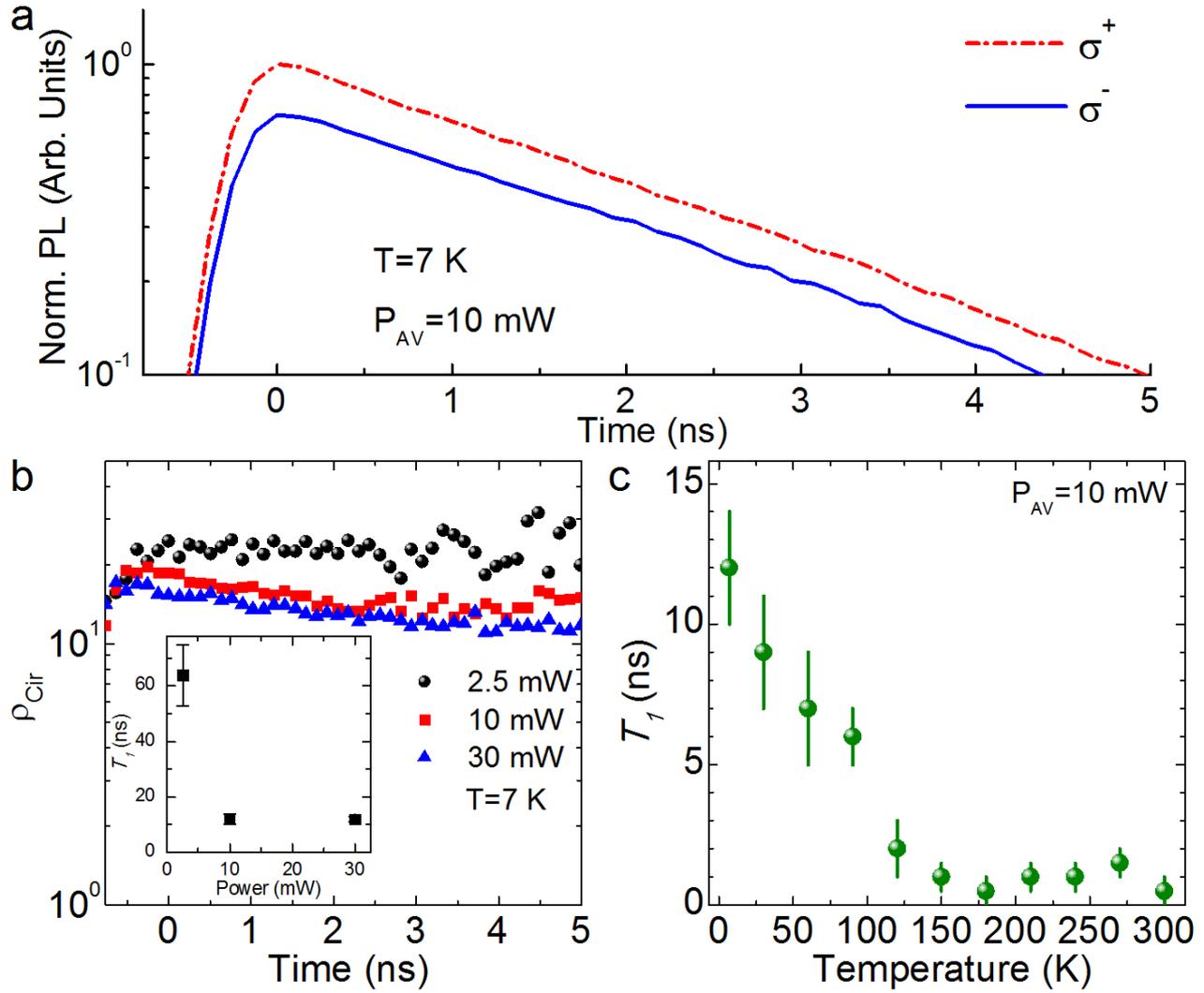

**Figure 4| Spin relaxation time. a** Temporal evolution of the polarization-resolved PL of $Ge_{0.95}Sn_{0.05}$ at an average pump power, $P_{AV}$, of 10 mW and a lattice temperature of 7 K. Right-handed circular polarization ($\sigma^+$) is shown as a dashed red line, whereas the left-handed helicity ($\sigma^-$) is reported as a blue line (the excitation laser is $\sigma^+$ polarised). **b** Low-temperature decay dynamics of the circular polarization degree $\rho_{circ} = \frac{I^{\sigma^+} - I^{\sigma^-}}{I^{\sigma^+} + I^{\sigma^-}}$, where $I^{\sigma^+}$ and $I^{\sigma^-}$ are the PL circular components. Black dots, red squares and blue triangles refer to an average pump power equal to 2.5, 10 and 30 mW, respectively. The inset demonstrates the derived spin relaxation time, $T_1$, as a function of the excitation power. **c** Temperature dependence of $T_1$ for $P = 10$ mW.

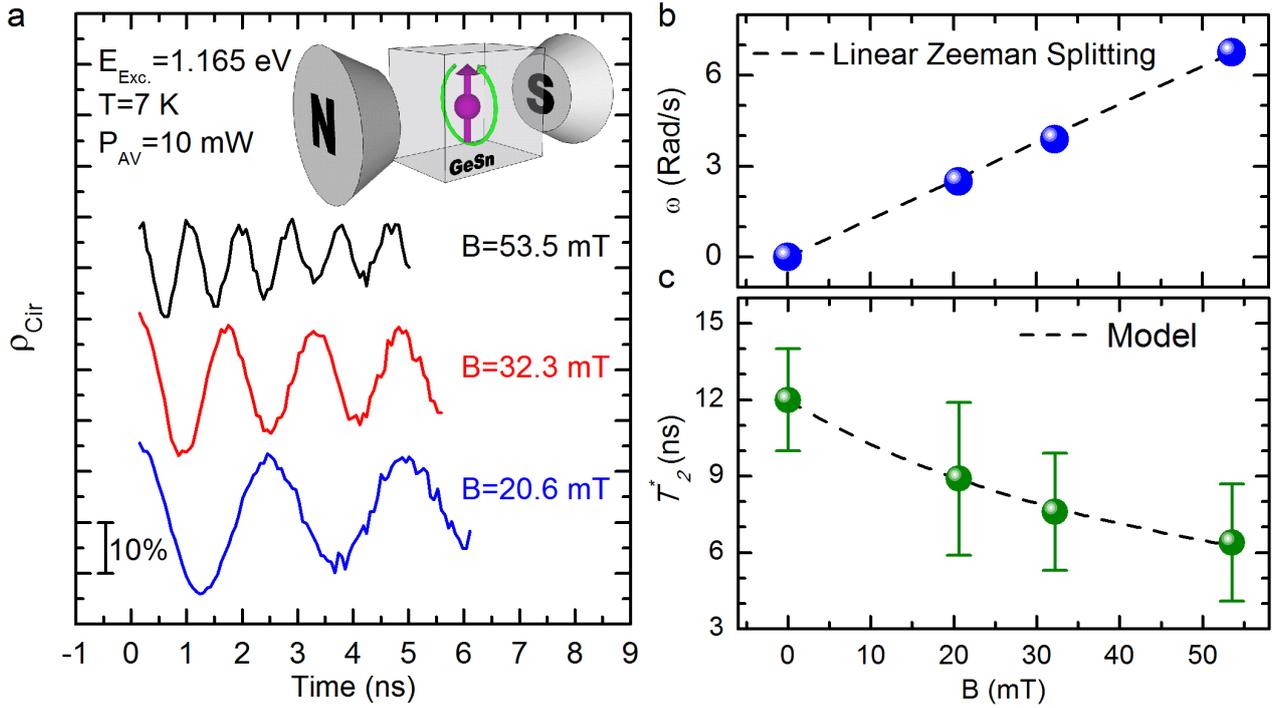

**Figure 5| Spin quantum beats in strained GeSn. a** Magnetic field dependence of the circular polarisation degree ($\rho_{circ}$) dynamics measured at a lattice temperature of 7 K and for a laser excitation energy of 1.165 eV and $P_{AV}$ =10 mW. Data obtained by applying a magnetic field strength of 20.6 (blue curve), 32.3 (red curve) and 53.5 mT (black curve) have been shifted for clarity. $\rho_{circ}$ oscillates around zero and the quantum beatings can be interpreted in terms of the Larmor precession of the electron spins due to the presence of an external magnetic field perpendicular to the quantization axis (Voigt configuration, see inset). **b** Dependence of the Larmor frequency, ω, on the magnetic field strength for a fully strained $Ge_{0.95}Sn_{0.05}$ layer epitaxially deposited on a Ge buffered Si substrate. The dashed line corresponds to the linear dependence determined by the Zeeman splitting $\hbar\omega = g^*\mu_B B$, where $g^*$ is the effective $g$-factor and $\mu_B$ is the Bohr's magneton. **c** Ensemble spin coherence time $T_2^*$ as a function of the external magnetic field strength. Experimental data have been obtained by the magnetic field dependence of the damping of the quantum beats. The dashed line in the figure is based on a modelling that correlates the spin dephasing time and the microscopic $g$-factor spread in the alloy layer.

24